\documentclass{osa-article}

\usepackage[obeyDraft]{todonotes}
\usepackage{mathrsfs}
\usepackage{float}

\journal{osajournal}

\articletype{Research Article}

\begin{document}

\title{Optical Parametric Amplification of Mid-Infrared Few-Cycle Pulses}

\author{
    Adam S.\ Wyatt,\authormark{1,2,*} 
    Paloma Mat\'{i}a-Hernando,\authormark{3}
    Allan S.\ Johnson,\authormark{3}
    Danylo T.\ Matselyukh,\authormark{1}
    Alfred J.\,H.\ Jones,\authormark{1}
    Richard T.\ Chapman,\authormark{1}
    Cephise Cacho,\authormark{1}
    Dane R.\ Austin,\authormark{3}
    John W.\,G.\ Tisch,\authormark{3} 
    Jon P.\ Marangos,\authormark{3} and 
    Emma Springate,\authormark{1}
}

\address{
    \authormark{1}Central Laser Facility, STFC Rutherford Appleton Laboratory, Harwell OX11 0QX, UK\\
    \authormark{2}Clarendon Laboratory, Department of Physics, University of Oxford, Oxford OX1 3PU, UK\\
    \authormark{3}Blackett Laboratory, Imperial College London, SW7 2AZ, UK
}
\email{\authormark{*}adam.wyatt@stfc.ac.uk} 



\begin{abstract}
We describe Ti:Sapphire pumped optical parametric amplification of carrier-envelope phase stabilized few-cycle ($<10$\,fs) mid-infrared pulses in type I $\beta$-barium borate. Experimental measurements show a $\times3.5$ amplification factor (from 100\,\textmu{}J to 350\,\textmu{}J) of the octave spanning spectrum ($1.1-2.4$\,\textmu{}m) using a pump beam with 2.3\,mJ energy, 30\,fs duration and central wavelength of 800\,nm, corresponding to an energy extraction efficiency of $11\%$. Numerical simulations suggest potential amplification to 4.25\,mJ energy and temporal compression to a pulse duration of 7.3\,fs is possible with a pump energy of 30\,mJ and duration of 30\,fs in a 25\,mm diameter, 1.5\,mm thick BBO crystal.
\end{abstract}

\section{Introduction}
\label{sec:Introduction}

Millijoule-level few-cycle (FC) carrier envelope phase (CEP) stabilized laser pulses in the mid-infrared (MIR) spectral region, with central wavelength $\sim2$\,\textmu{}m \cite{Austin2016, Fan2016a}, have recently proved to be a valuable tool in ultrafast science since they can be used in strong field experiments \cite{Wolter2015} and for the generation of high flux coherent radiation confined to sub-femtosecond duration and with a continuous spectrum spanning the water window ($\sim280-530$\,eV) via the process of high harmonic generation (HHG) \cite{Popmintchev2012a, Cousin2014, Ishii2014, Silva2015, Teichmann2016, Li2017, Johnson2018}. This soft x-ray spectral region is especially important due 
to the presence of many absorption edges of chemically and biologically important species such as carbon, nitrogen, oxygen (K-edges at 284\,eV, 410\,eV and 530\,eV respectively) and sulfur (L-edges ranging from 164--229\,eV) \cite{Kasrai1990, Lafuerza2011}. When combined with the attosecond temporal resolution afforded by the HHG process with a FC pulse \cite{Chipperfield2005, Sansone2006, Goulielmakis2008, Chini2014}, the measurement of transient x-ray absorption near-edge structure (XANES) with sufficient temporal resolution to track electron and hole dynamics of excited state molecules and surfaces becomes a real possibility \cite{Johnson2016, Pertot2017, Young2018}. 

HHG \cite{Burnett1977, McPherson1987, Ferray1988, LHuillier1989, Schafer-1997-High, frank2012invited} in gases is described by a simple three step model in which an electron tunnel ionizes in a strong laser field, is accelerated away from and then driven back into the parent ion, emitting an attosecond burst of extreme ultraviolet (XUV) radiation every half cycle \cite{Corkum1993, Lewenstein1994}. Using a longer wavelength increases the ponderomotive energy for a given laser intensity, and thus can generate higher energy photons whilst minimizing free electrons from ionization, which is important to ensure the HHG process is phase-matched. However, a longer wavelength also significantly reduces the total flux. Additionally, a FC driver reduces ionization of the leading edge of the pulse enabling transient phase-matching to occur as well as restricting generation of the maximum photon energies to a single attosecond burst, yielding a broad continuous spectrum. For example, HHG driven by a 40\,fs, 1.5\,mJ, 800\,nm pulse at a repetition rate of 1\,kHz resulted in an XUV flux of $10^{10}$\,ph/s/harmonic at 23\,eV \cite{Constant1999} whereas HHG driven by FC MIR pulses at $\sim2$\,\textmu{}m have generated a soft x-ray flux of $10^8-10^9$\,ph/s over the whole water window spectrum (300--600\,eV) \cite{Teichmann2016, Johnson2018}. Therefore it is important to use the shortest driving wavelength possible that phase-matches HHG at the desired photon energy; using a FC driving pulse with a wavelength of $\sim1.8$\,\textmu{m} turns out to be ideal for generating water window harmonics.

In this paper, we propose an optical parametric amplifier (OPA) \cite{Ross-2002-Analysis, cerullo2003ultrafast} capable of amplifying FC MIR CEP stable pulses from the sub-millijoule energy level to multiple millijoules and verify the scheme through numerical simulations and experimental measurements. This scheme is simple to implement with a moderately small investment and will lead to an increase in yield of water-window harmonics as well as allowing access to relativistic pulse intensities ($\sim10^{18}$\,W/cm$^2$) within the MIR spectral region.

\subsection{Mid-Infrared Few-Cycle Pulse Generation}

A commonly employed method of generating MIR ultrafast pulses is to use an OPA pumped using either Ti:Sapph ($\sim800$\,nm) \cite{Vozzi2007} or ytterbium ($\sim1030$\,nm) based lasers \cite{Fernandez2009}. Ti:Sapph lasers are a mature technology, already commonplace in ultrafast laser labs and due to favourable phase matching in BBO can produce $\sim1$\,mJ pulses with $\sim15\%$ efficiency and $\sim1.5$ times the pump duration using commercially available OPAs. In such an OPA, the $\sim1.5$\,\textmu{}m signal beam is generated by white-light generation (WLG) in sapphire using a small fraction of the pump energy, resulting in an idler centred at $\sim1.7$\,\textmu{}m. Since both the pump and signal are derived from the same laser, the idler is passively CEP stable \cite{Cerullo2011}. Additional active stabilization is typically employed by adjusting the pump--signal delay to compensate for intensity noise and thermal drift. This output can then be compressed to the few-cycle regime using self-phase modulation (SPM) inside a gas-filled hollow core fiber (HCF), generating a spectrum spanning $\sim1.1-2.4$\,\textmu{}m. Fused silica exhibits anomalous dispersion above 1.3\,\textmu{}m and can therefore be used to compensate the normal dispersion introduced by the gas dispersion and SPM, as well as controlling the CEP. 

There are several issues with scaling the energy from external pulse compression in gas filled capillaries: (1) increasing the input pulse energy up to 10\,mJ \cite{Thire2015} is possible by increasing the capillary inner diameter, but a longer fibre is also required, e.\,g.\ on the order of 3\,m, making the compression setup impractical, especially for small-scale laser laboratories; (2) the external pulse compression net efficiency is on the order of 50\%, depending on the spatial mode quality, and thus much of the pump energy converted into the MIR is not transferred into the FC pulse; and (3) higher order spectral phase resulting from the SPM and post dispersion compensation in fused silica limit the compressibility of the pulse to $\lesssim10$\,fs with a non-insignificant fraction of energy in temporal side-bands/pedestals --- generating a pulse duration of $\sim7$\,fs is possible but with increased energy in the temporal side-bands and thus not significantly increasing the peak intensity or reducing pre-ionization.

\begin{figure}[H]
    \centering\includegraphics[width=\textwidth]{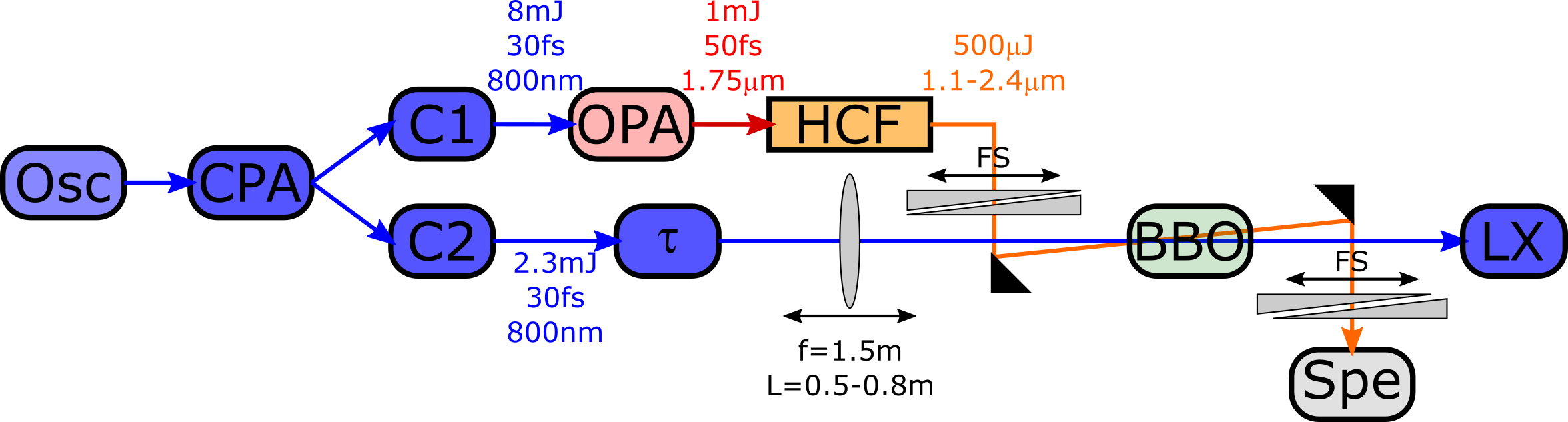}
    \caption{Schematic of the experimental setup. Osc: Ti:Sapph oscillator; CPA: Ti:Sapph chirped pulse amplifier; C1/C2: grating comressor; OPA: optical parametric amplifier (Light Conversion HE-TOPAS-Prime); HCF: hollow core compressor; FS: fused silica wedges; $\tau$: delay; f: focal length; L: distance from lens to BBO crystal; BBO: $\beta$-barium borate nonlinear crystal; LX: LX-SPIDER pulse characterization; Spe: MIR Spectrometer (Ocean Optics NI-Quest). The output of the CPA is split into two beam lines that are compressed independently; the high energy beam line is used to pump a commercial OPA, the idler of which is spectrally broadened in a HCF and dispersion tuned using a pair of FS wedges, whilst the lower energy line is delayed and loosely focused beyond the BBO crystal to be used as a pump for our FC MIR OPA. The pump pulse is measured using an LX-SPIDER and the amplifier FC MIR spectrum is measured using a fibre-coupled grating spectrometer. The position of the lens is adjusted to vary the pump beam size in the OPA and hence the pump peak intensity.}
    \label{fig:ExperimentalSetup}
\end{figure}

Here we propose a slightly alternative method of generating multi-millijoule energy few-cycle pulses in the MIR based on a few-cycle OPA, as illustrated in fig.~\ref{fig:ExperimentalSetup}. A relatively low energy few-cycle MIR pulse is generated, via external pulse compression inside a gas filled capillary, from the $\sim1.8$\,\textmu{}m idler exiting a Ti:Sapph pumped commercial OPA. This FC pulse is then amplified further in a single stage 800\,nm pumped type I BBO OPA. Due to the high pulse intensities, large spectral bandwidth and required low dispersion, practical difficulties in collinearly overlapping the pump and seed beams meant we used a slightly non-collinear geometry with an external angle between the pump and seed of $\sim2^\circ$. Note that our modelling suggests a collinear geometry is also feasible. This could be achieved for example using an s-polarized Fresnel reflection for the seed and p-polarized transmission at Brewster angle for the pump. The residual transmitted ``seed'' beam could then be utilized as a ``pump'' pulse for performing pump-probe spectroscopy.

\section{Numerical Modelling}
\label{sec:Simulations}

In order to investigate the potential OPA performance for short pulses, we numerically resolve the forward-propagating first order coupled wave equations~(\ref{eq:WaveEquation}) using a split-step Runge-Kutta method in the plane-wave approximation limit (i.\,e.\ only in the longitudinal spatial dimension and thus neglecting e.\,g.\ diffraction and spatial walk-off),
\begin{equation}
    \label{eq:WaveEquation}
    \partial_z\widetilde{U}_n(\omega, z) \approxeq \frac{2i\omega d_\text{eff}}{n(\omega)c}\widetilde{p}_n(\omega, z)e^{-i k(\omega)z},
\end{equation}
where $\widetilde{p}_n(\omega, z)$ are the nonlinear polarization response functions for the pump (p), signal (s) and idler (i) pulses given by
\begin{align}
    \label{eq:NoncollinearPolarization}
    \widetilde{p}_\text{p}(\omega, z) &= \mathscr{F}\left\{ E_\text{s}E_\text{i}(t, z) \right\}\\\nonumber
    \widetilde{p}_\text{s}(\omega, z) &= \mathscr{F}\left\{ E_\text{p}(t,z)E_\text{i}^\ast(t, z) \right\}\\\nonumber
    \widetilde{p}_\text{i}(\omega, z) &= \mathscr{F}\left\{ E_\text{p}(t,z)E_\text{s}^\ast(t, z) \right\}. 
\end{align}
and $E_n(t,z) = \mathscr{F}^{-1}\left\{\widetilde{U}_n(\omega,z)e^{ik_n(\omega)z}\right\}$ are the analytic signals of the interacting electric fields such that the real electric fields are $\varepsilon_n = 2\mathscr{R}\left\{E(t,z)\right\}$ and the intensities are $I_n(t, 0) = 2\epsilon_0 c | \mathscr{F}^{-1} \{ \sqrt{n(\omega)} \widetilde{U}_n(\omega, 0) \} |^2$; $\mathscr{F}\{\}$ and $\mathscr{R}\{\}$ represent the Fourier transform and real operators; $k_n(\omega) = [n(\omega)\omega/c]\cos\alpha_n$ and $z$ is the longitudinal wave-vector and distance respectively in the pump direction; $\alpha_n$ is the internal angle of the n$^\text{th}$ beam relative to the pump; $n(\omega)$ is the refractive index at frequency $\omega$; $d_\text{eff}$ is the effective non-linearity for the crystal and geometry; and $c$ is the speed of light in vacuum.

\begin{figure}[H]
    \centering\includegraphics[width=\textwidth]{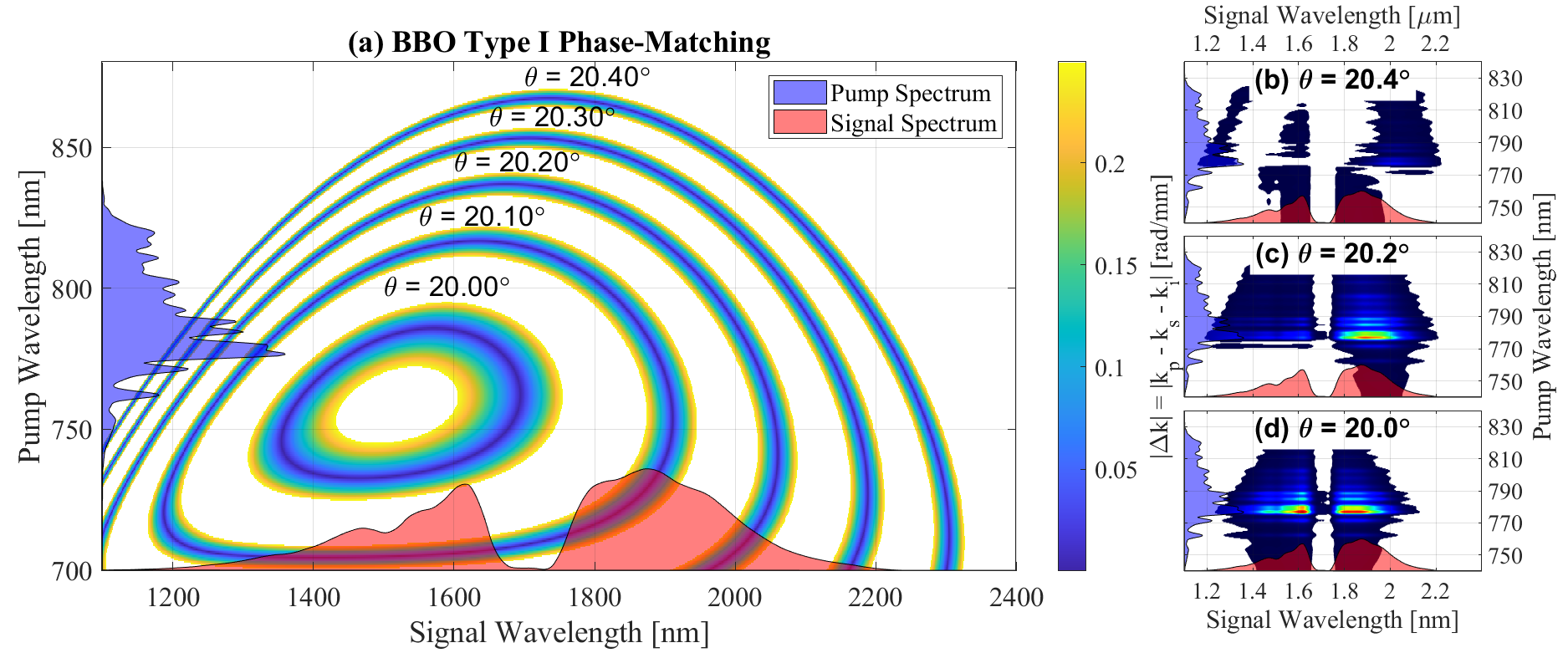}
    \caption{(a) Phase-matching curves for Type I DFG in BBO for various phase-matching angles, $\theta$ of the crystal. (b)-(d) OPA CW-gain for each pump and signal wavelength. Shaded curves: measured (blue) pump and (red) seed spectral intensities.}
    \label{fig:PhaseMatching}
\end{figure}

\begin{figure}[H]
    \centering\includegraphics[width=\textwidth]{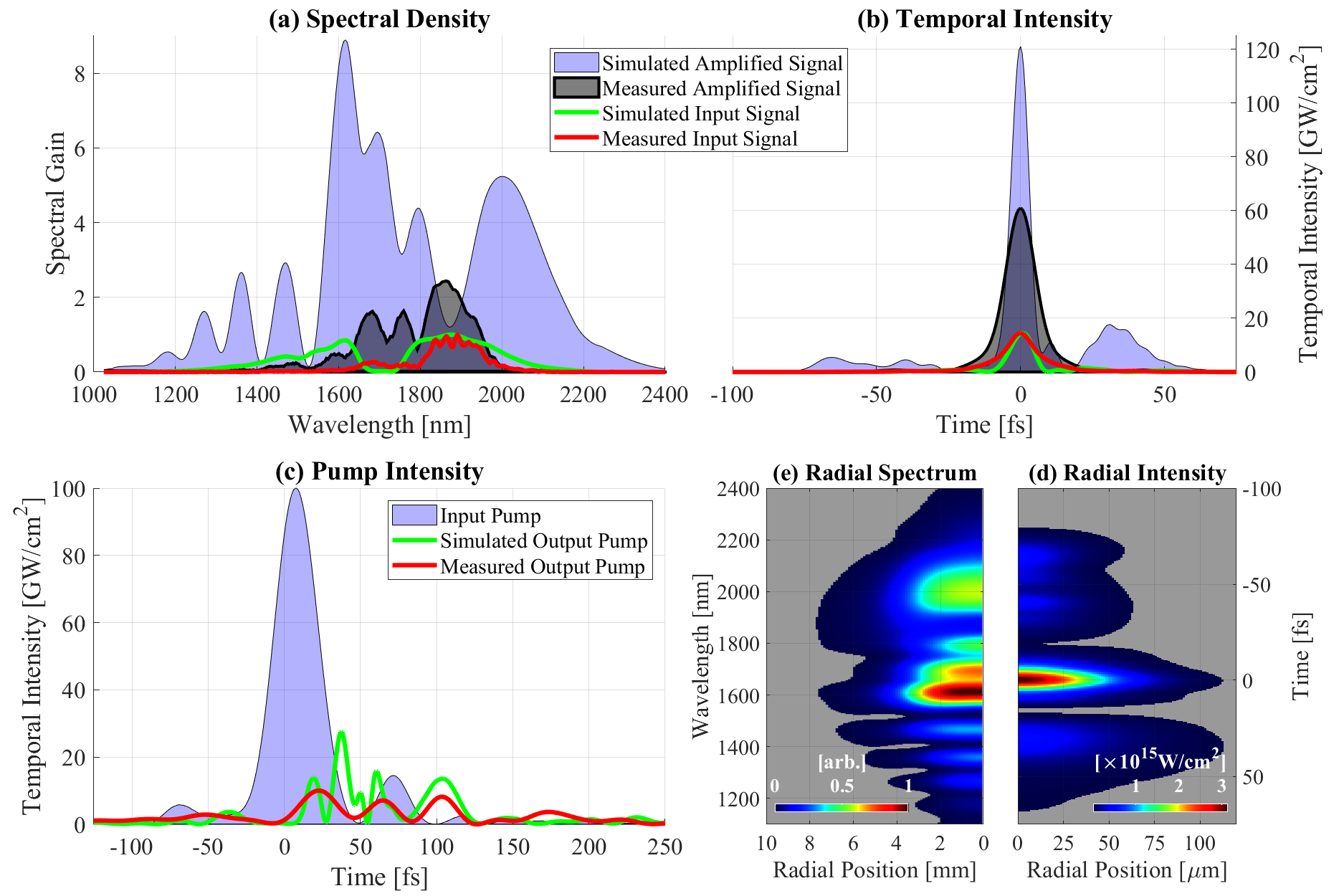}
    \caption{(a) Measured and simulated seed spectrum before and after the BBO crystal. (b) Simulated temporal intensity plus the Fourier transform limited temporal intensity resulting from the measured spectrum of the seed before and after the BBO. (c) Measured temporal intensity of the pump scaled to 100\,GW/cm$^2$ peak intensity used in both the OPA measurements and simulations, plus the measured and simulated temporal intensity of the pump after the OPA crystal, showing the strong depletion and temporal oscillations. (d) Radial dependence of the simulated amplified seed spectrum. (e) Radial dependence of the temporal intensity of the simulated amplified seed at the focus of a -1\,m radius of curvature spherical mirror.}
    \label{fig:Gain}
\end{figure}

The choice of nonlinear crystal was chosen by considering the continuous-wave (CW) phase-matching conditions for all difference frequency generation (DFG) combinations of the pump and signal wavelengths. The resulting phase-matching curves for type I $\beta$-barium borate (BBO) for a range of phase matching angles is plotted in fig.~\ref{fig:PhaseMatching} along with the corresponding CW gain in the limit of zero pump depletion, indicating a collinear phase-matching of $\sim20^\circ$ is optimal. We then integrated equation~(\ref{eq:WaveEquation}) over a crystal length of 1.5\,mm. For the simulations, we used an LX-SPIDER to provide the temporal intensity and phase for the Ti:Sapph pump pulse --- see fig.~\ref{fig:Gain}(b); for the MIR seed we used we used the measured pulses exiting the HCF in \cite{Austin2016}. The pump pulse had an energy of 2.3\,mJ and a duration of 33\,fs full width at half maximum (FWHM), which (assuming a Gaussian spatial profile) requires a $e^{-2}$ radial beam width of 5.8\,mm to obtain a peak intensity of 100\,GW/cm$^2$. The seed pulse had a FWHM pulse duration of 15.7\,fs at the exit of the HCF and an optimal FWHM pulse duration of 9.3\,fs after dispersion from 1.9\,mm of fused silica. Assuming 100\,\textmu{}J input pulse energy and the same beam radius as the pump, the peak intensity of the seed is 5.7\,GW/cm$^2$ and 14.5\,GW/cm$^2$ at the exit of the HCF and optimally compressed respectively. 

We simulated a global optimization routine in which we varied the dispersion from fused silica inserted into the signal beam before the BBO, the crystal phase-matching angle, and the non-collinear angle and delay between pump and signal beams. We kept the following parameters fixed: the crystal thickness at 1.5\,mm, pump pulse intensity at 100\,GW/cm$^2$, pump dispersion as measured using the LX-SPIDER and the MIR seed fluence at 186\,\textmu{}J/cm$^2$ (100\,\textmu{}J energy, 5.8\,mm waist). We then simulated additional dispersion from fused silica after the BBO to compress the pulse to yield the highest peak intensity for a given set of optimization parameters. The optimum parameters corresponded to 3.2\,mm of fused silica before amplification, $21.44^\circ$ phase-matching angle, $3^\circ$ internal non-collinear angle and 50\,fs delayed pump. Although energy is predominantly transferred from the pump into the signal and idler, some energy is transferred back into the pump as the pulses temporally walk-off from each other due to group velocity mis-match. As a result, the depleted pump pulse exiting the BBO exhibits oscillations in its temporal intensity, as shown in fig.~\ref{fig:Gain}(c). The simulated MIR spectra before and after the BBO are plotted in fig.~\ref{fig:Gain}(a) and the corresponding temporal intensities in fig.~\ref{fig:Gain}(b). The simulated output seed fluence was 1.55\,mJ/cm$^2$, resulting in a fluence gain factor of $\times8.3$. After applying the optimal amount of fused silica after the BBO, the simulated amplified seed has a peak intensity of 121\,GW/cm$^2$ and FWHM pulse duration of 7.28\,fs; corresponding to an intensity gain factor of $\times8.3$ and pulse compression factor of $\times0.78$. We then independently simulated the amplified signal with the optimal parameters for each radial position in the beam; the output spectrum is plotted as a function of radial co-ordinate in fig.~\ref{fig:Gain}(d) and the temporal intensity of the beam at the focus of a spherical mirror with radius of curvature of -1\,m is plotted in fig.~\ref{fig:Gain}(e). The simulated energy of the amplified beam was 520\,\textmu{}J, corresponding to an energy gain factor of $\times5.2$ and relatively constant pulse duration and group delay across the spatial profile.

\begin{figure}[H]
    \centering\includegraphics[width=\textwidth]{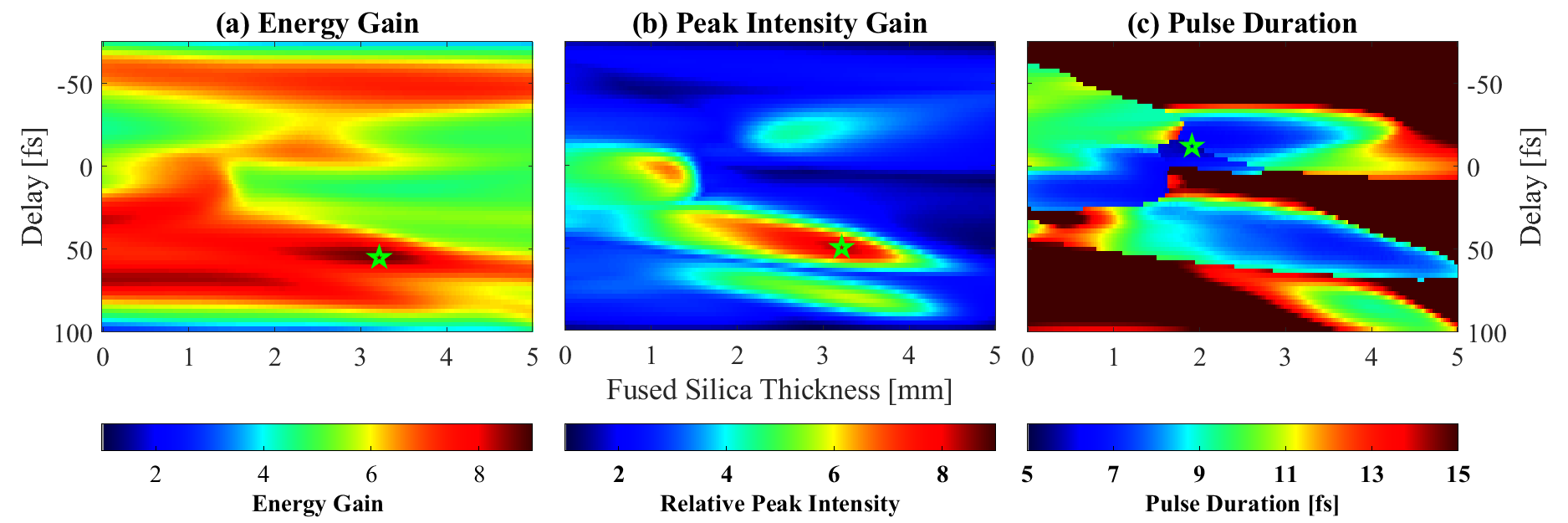}
    \caption{Simulated parameter dependence on pump delay and seed dispersion from the OPA (see main text for details).}
    \label{fig:SimulatedParameterScan}
\end{figure}

Another important aspect to consider is the performance of the OPA as a function of the input parameters. We therefore simulated the OPA using the same parameters as above, but over a range of pump-signal delay and fused silica dispersion before the BBO, followed by a fixed amount of fused silica after the BBO. We then considered three different parameters, energy gain (assuming top-hat spatial profile), peak intensity gain (relative to the fully compressed pulse without amplification) and FWHM pulse duration --- the results are shown in fig.~\ref{fig:SimulatedParameterScan}. Since the performance varies smoothly around the optimum parameters (marked by the green star), we can conclude that we expect the OPA to not be sensitive to small drifts and fluctuations. Note also that there are two regions showing good performance: (1) near zero delay and 2\,mm of fused silica; and (2) near 50\,fs pump delay and 3\,mm of fused silica. Both BBO and fused silica have the same sign third order phase, thus minimizing the material dispersion results in a shorter pulse duration, corresponding to the first optimum. However, pre-delaying the pump pulse enables the signal to temporally walk-through the pump pulse, maximizing energy extraction efficiency and leading to higher output energy, corresponding to the second optimum. It is therefore important to consider what performance parameter is most important to the experiment and optimize on that. For example, the optimum peak intensity will result through a combination of efficient energy extraction as well as a short compressed pulse, and so this is a good candidate to consider for HHG and strong-field experiments. 

\section{Experimental Results}
\label{sec:ExperimentalResults}

Using the setup depicted in fig.~\ref{fig:ExperimentalSetup}, we performed proof-of-principle experiments to verify the performance of the proposed OPA. In order to vary the pump intensity, a lens was used to loosely focus the pump some distance beyond the crystal. The lens was positioned on a motorized translation stage in order to move the distance of the waist relative to the crystal, therefore changing the pump mode size on the crystal. A fiber-coupled infrared spectrometer was used to measure the spectrum of the signal at the center of the beam when the pump is either blocked or unblocked. A set of SF10 wedges instead of fused silica were used to control the seed dispersion before the BBO crystal. The lens position, pump delay and SF10 wedge insertion were all computer controlled. A power meter placed after an iris was used to measure the beam intensities. The position of the lens was chosen as a compromise between good spatial overlap with the seed beam profile and maximizing the pump intensity, but keeping below the damage threshold of the crystal. We estimate the pump intensity to be 100\,GW/cm$^2$.

The measured MIR spectrum before and after amplification are plotted in fig.~\ref{fig:Gain}(a), and the corresponding Fourier transform limited intensity resulting from these spectra are plotted in fig.~\ref{fig:Gain}(b) (at the time of the experiment, we had no device capable of measuring the spectral phase or temporal intensity of the FC MIR pulse). The measured temporal intensity of the pump before and after the OPA is plotted in fig.~\ref{fig:Gain}(c). Note that the measured and simulated pump temporal intensity profiles exiting the BBO show good agreement with each other are similar even though the exact signal pulses differ. We measured a gain factor of $\times3.5$, resulting in an amplified seed energy of 350\,\textmu{}J from an input energy of 100\,\textmu{}J (the part of the seed beam that overlapped with the pump beam), corresponding to an energy extraction efficiency of $\sim11\%$. Another important feature to note is that the spectrum in the wings are amplified by a larger factor than in the centre, effectively resulting in a larger bandwidth. Increasing the pump intensity by reducing the pump beam radius, it was possible to obtain even more gain, over $\times7.5$, closer to the simulated values. In order to prevent damage to the BBO crystal due to intensity fluctuations and hot-spots in the pump beam spatial profile, the remaining measurements were performed at a more conservative pump intensity.

\begin{figure}[H]
    \centering\includegraphics[width=\textwidth]{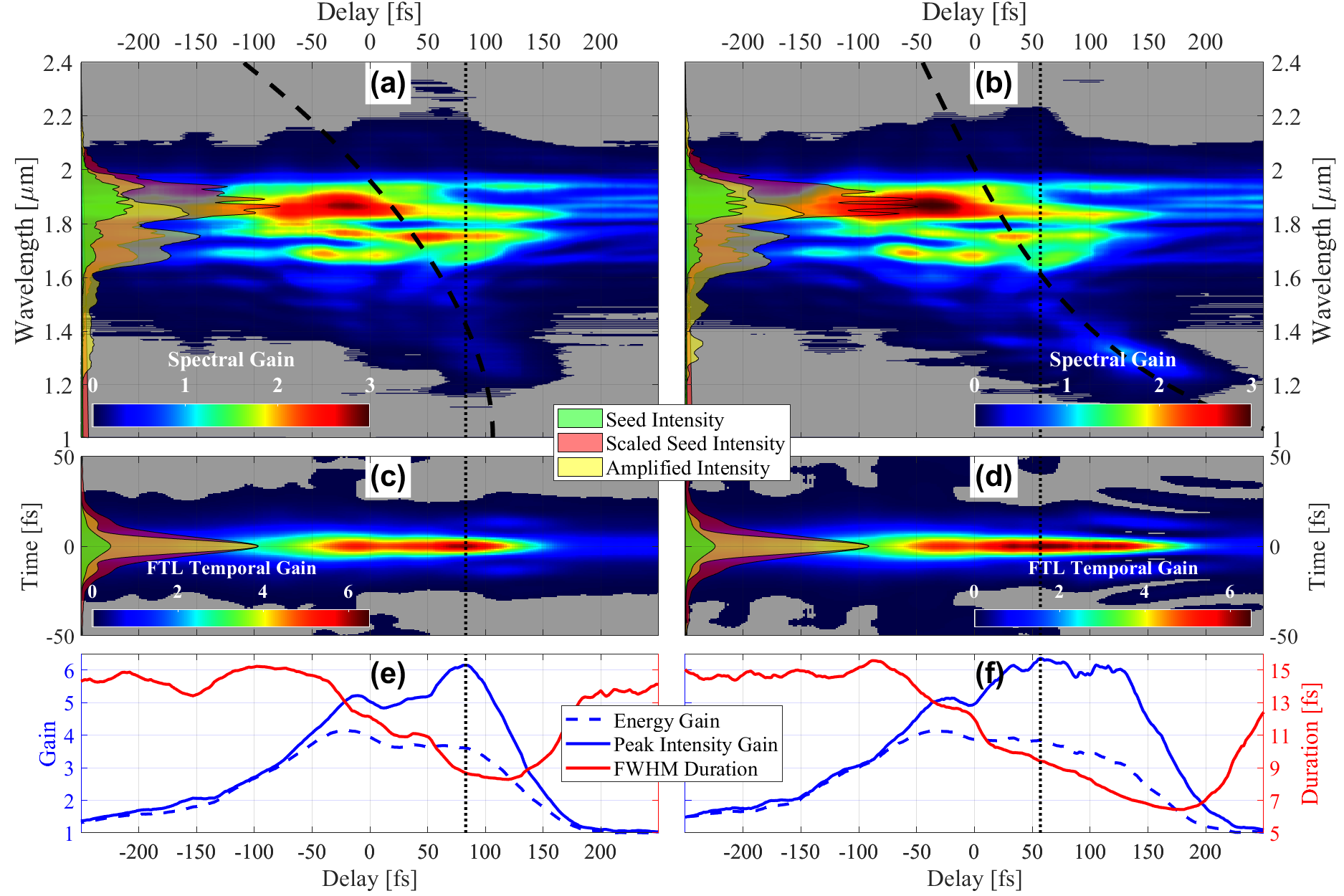}
    \caption{Measured OPA dependence on pump delay for two values of pre-dispersion: SF10 thickness of (left column) 0.94\,mm and (right column) 4.4\,mm. (a,b) Measured amplified MIR spectrum; dashed black line indicates wavelength dependent group delay of the SF10 wedges minus 5\,mm fused silica (corresponding to the approximate amount of dispersion required to compress the output of the HCF); dotted yellow line marks the ``optimal'' delay resulting in the maximum Fourier transform limited (FTL) peak intensity; shaded curves on left correspond to the (green) seed spectrum, (yellow) amplified spectrum at optimum delay and (red) seed spectrum scaled to the same energy as the amplified spectrum. (c,d) FTL temporal intensity (i.e. Fourier transform of spectral amplitude in (a,b) assuming no spectral phase); shaded regions on left indicate FTL temporal intensities of (green) seed, (yellow) amplified seed at optimum delay and (red) seed scaled to same peak intensity as amplified seed at optimum delay. (e,f) Extracted parameter on delay: (dashed blue) energy gain and (solid blue) FTL peak intensity gain using left-hand scale, (red) full width at half maximum (FWHM) duration of FTL intensity using right-hand scale.}
    \label{fig:MeasuredDelayDependence}
\end{figure}

We then performed a scans of the amplified seed spectrum as a function of the dispersion from SF10 wedges introduced to the seed before the BBO and as a function of the pump delay. The pump-delay scan results for two different SF10 wedge insertion settings are plotted in fig.~\ref{fig:MeasuredDelayDependence} along with extracted pulse parameters. It is clear that the dispersion from the SF10 is mapped onto the delay dependence of the spectrum: large SF10 dispersion results in the shorter wavelengths arriving at the trailing edge of the pulse and thus being amplified at larger pump delays. The variation in the amplified bandwidth and thus FTL pulse duration as a function of delay is quite clear, especially when we consider the performance over all delays and SF10 wedge positions, as shown in fig.~\ref{fig:SimulatedParameterScan}. Even though experimentally the maximum energy extracted did not strongly depend on the SF10 wedge thickness, the spectral bandwidth and hence FTL pulse duration depends quite strongly on the seed dispersion. Although a direct measurement of the MIR spectral phase would be required, we can see that the parameters that yield the optimum bandwidth do not necessarily match those that optimize the energy, similar to what was observed in the simulated data.

\begin{figure}[H]
    \centering\includegraphics[width=\textwidth]{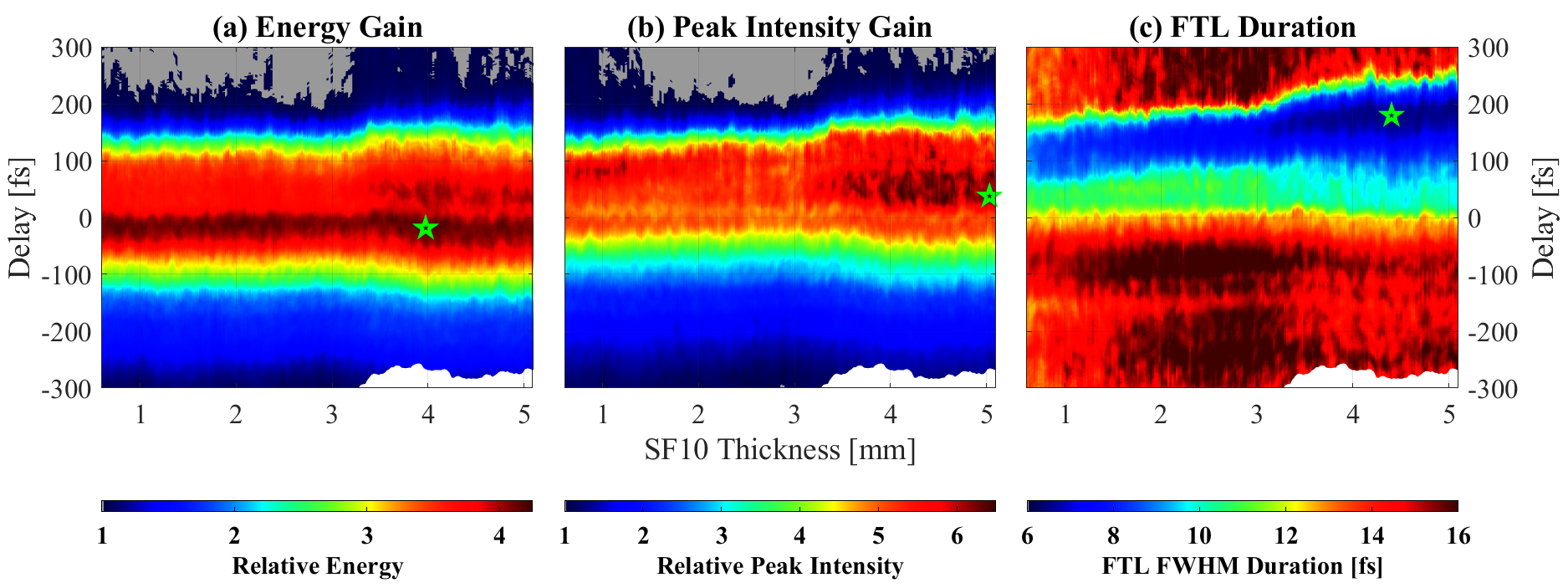}
    \caption{Measured OPA parameter dependence on pump delay and SF10 pre-dispersion. (a) Energy gain (amplified spectral energy / input spectral energy). (b) Peak intensity gain (amplified FTL peak intensity / input FTL peak intensity). (c) FTL pulse duration (FWHM). In each plot, the global maximum gain and minimum duration are marked by the green stars.}
    \label{fig:MeasuredParameterScan}
\end{figure}

\section{Conclusion}
\label{sec:Conclusion}

We have experimentally demonstrated amplification of the full octave-spanning spectrum of a CEP stable FC MIR pulse in a single stage type I BBO OPA pumped by a 33\,fs 2.3\,mJ Ti:Sapphire pulse centered at 800\,nm. We measured an amplification factor of $\times3.5$ (increasing to $\times7.5$ at higher pump intensity near the damage threshold) with $11$\% energy transfer from the pump to the seed. Although temporal characterization of the amplified seed is required, we observed that the spectral bandwidth was also increased, indicating potential compression in the pulse duration, in corroboration with simulated data. We are currently upgrading our laser system to be able to provide 10\,mJ pump energy for the OPA. Assuming a seed energy of $500$\,\textmu{}J, optimized mode matching between the pump and seed and $10$\% energy transfer from the pump, we estimate that we are able to generate MIR pulses with $>1.5$\,mJ energy, although we predict that with further optimization of the pump intensity and OPA parameters (e.\,g.\ pre-dispersion, delay and phase-matching angle) we can increase this further to closer match our numerical models. Since the maximum available aperture size of BBO is limited to 25\,mm, the maximum pump energy that can be used is limited to $\sim30$\,mJ, assuming a top-hat spatial profile and $30$\,fs duration (FWHM, Gaussian temporal profile), which is commercially available at 1\,kHz repetition rates, this would lead to an amplified seed energy of 4.25\,mJ with a pulse duration of 7.3\,fs and peak power of 330\,GW, sufficient to reach a peak intensity of $10^{18}$\,W/cm$^2$ with a sub-5\,\textmu{}m focal spot.


\section*{Funding}
Science and Technology Facilities Council (STFC); European Commission FP7 LASERLAB-EUROPE III (284464); EPSRC/DSTL MURI Grant EP/N018680/1.

\section*{Acknowledgments}
The authors would like to thank the technical support of Phil Rice, Dave Rose and the CLF mechanical engineering department.




\end{document}